\documentclass[prd,twocolumn,preprintnumbers]{revtex4}

\usepackage{psfrag}
\usepackage{graphicx}
\usepackage{graphics}
\usepackage{bm}
\usepackage{color}
\usepackage{amssymb}
\usepackage{amsmath}

\newcommand{\be}{\begin{equation}}
\newcommand{\ee}{\end{equation}}
\newcommand{\ba}{\begin{eqnarray}}
\newcommand{\ea}{\end{eqnarray}}

\newcommand{\nn}{\nonumber}

\newcommand{\gsim}{\raise.3ex\hbox{$>$\kern-.75em\lower1ex\hbox{$\sim$}}}
\newcommand{\lsim}{\raise.3ex\hbox{$<$\kern-.75em\lower1ex\hbox{$\sim$}}}

\begin{document}
\preprint{CERN-PH-TH/2008-056}

\renewcommand{\thefootnote}{\fnsymbol{footnote}}

%-------------------------------------------------

\renewcommand{\thefootnote}{\arabic{footnote}}
\setcounter{footnote}{0} \typeout{--- Main Text Start ---}

\title{Statistical mechanics of strings with Y-junctions}
\author{R.~J.~Rivers}
\email{r.rivers@imperial.ac.uk} \affiliation{Blackett Laboratory,
Imperial College, London SW7 2AZ, United Kingdom}
\author{D.~A.~Steer}
\email{steer@apc.univ-paris7.fr} \affiliation{APC, B\^atiment
Condorcet, 10 rue Alice Domon et L\'eonie Duquet 75205 Paris Cedex
13, France \\
CERN Physics Department, Theory Division, CH-1211 Geneva 23, Switzerland}

\date{\today}

\begin{abstract}

We investigate the Hagedorn transitions of string networks with
Y-junctions as may occur, for example, with ($p,q$) cosmic
superstrings.  In a simplified model with three different types of string, the partition
function reduces to three generalised coupled XY
models. We calculate the phase diagram and show that, as the
system is heated, the lightest strings first undergo the Hagedorn
transition despite the junctions. There is then a second, higher, critical temperature
above which infinite strings of all tensions, and junctions, exist. Conversely,
on cooling to low
temperatures, only the lightest strings remain, but they
collapse into small loops.

\end{abstract}
\maketitle

\noindent

\section{Introduction}

The statistical mechanics of string networks has been the object of numerous studies because
of the importance of strings or string-like entities across all energy scales.

In general, either because of the large number of configurational
microstates or because of the large number of excited quantum
states that such a network possesses, the networks undergo
transitions in which, as temperatures rise, strings proliferate.
In the language of configurational states such a transition is
termed a Feynman-Shockley transition, after Feynman's description
of the $\lambda$-transition of $^4$He in terms of vortex
production \cite{Kleinert}. From the viewpoint of counting excited
states it is called a Hagedorn transition \cite{Hagedorn}.
[Henceforth we follow the common usage of {\it Hagedorn
transition} to apply to both cases, which are similar in structure
in many ways.]

Specifically, in QCD, the sudden proliferation of colour flux
tubes (the original dual hadronic strings) explains quark
deconfinement as temperature rises (see, for example,
\cite{Patel1,Patel2,MR}). In cosmology at the GUT scale, where
cosmic strings arise in all reasonable supersymmetric models
incorporating electroweak unification \cite{Mairi}, the
statistical mechanics of cosmic string networks has been
investigated in order to understand their properties at formation
and their late time scaling solutions, crucial for determining
their cosmological consequences \cite{EdRay2,EdRay3}. For
fundamental strings there has been substantial work on exploring
the effects of such transitions on the extremely early universe
\cite{AtickWitten,Mitchell1,Mitchell2,AlbrechtTurok,
Sakellariadou3}.

More recently, attention has turned again to fundamental string
networks, following new developments in superstring theory.
Indeed, a network of cosmic superstrings is expected to form when
a brane and anti-brane annihilate at the end of string-motivated
brane inflation models. The network contains fundamental F-strings, Dirichlet D-strings,
and $(p,q)$-strings which are bound states of $p$ F-strings and
$q$ D-strings
\cite{Copeland,m2,m3,Firouzjahi:2006vp}, meeting
at Y-junctions (or vertices).
The presence of Y-junctions, as well as the spectrum of tensions of the strings,
is a key characteristic of such
networks and leads to more complicated dynamics. Much work has
been done to determine how $(p,q)$-like string networks evolve, both by analytic methods
and numerical simulations, with particular regard to scaling
solutions, their effect on the CMB as well as other observable
consequences
\cite{Tye:2005fn,Sakellariadou:2004wq,Avgoustidis:2005nv,Copeland:2005cy,Saffin:2005cs,
Hindmarsh:2006qn,us1,Wells,Hassan,Jon,Arrtu,Siemens}.

Other than being stable against break-up, such strings differ from
earlier superstrings in that, due to the warping of space-time,
their tensions are not of the Planck scale but many orders of
magnitude smaller. As a result any Hagedorn transitions may even
arise later than the reheating of the universe, and hence be of
direct relevance for astrophysics. A necessary first step in
seeing whether this is the case is to determine the phase diagram
for the Hagedorn transitions of a  network with more than one type
of string, and this is the goal of the present paper.

Our approach is to attempt to map the thermodynamics of string
networks with junctions into the thermodynamics of a set of
interacting dual fields, whereby the Hagedorn transitions of the
strings become conventional transitions of the fields, a situation
with which we are familiar. One can imagine several ways to
attempt this. We adopt the simplest, generalising the methods for
describing quark deconfinement mentioned above (with its flux-tube
Y-junctions) to something more like $(p,q)$-strings.

Hence we investigate the {\it equilibrium
statistical mechanics} of cosmic superstring networks using
methods motivated by \cite{Patel1,Patel2,MR}. However, it
is important to note that there is at least one major difference between
cosmic superstrings and QCD fluxlines: with multiple tensions
(from different string types), we expect cosmic superstring
networks to show multiple Hagedorn transitions.

\begin{figure}
\centerline{\includegraphics[width=0.42\textwidth]{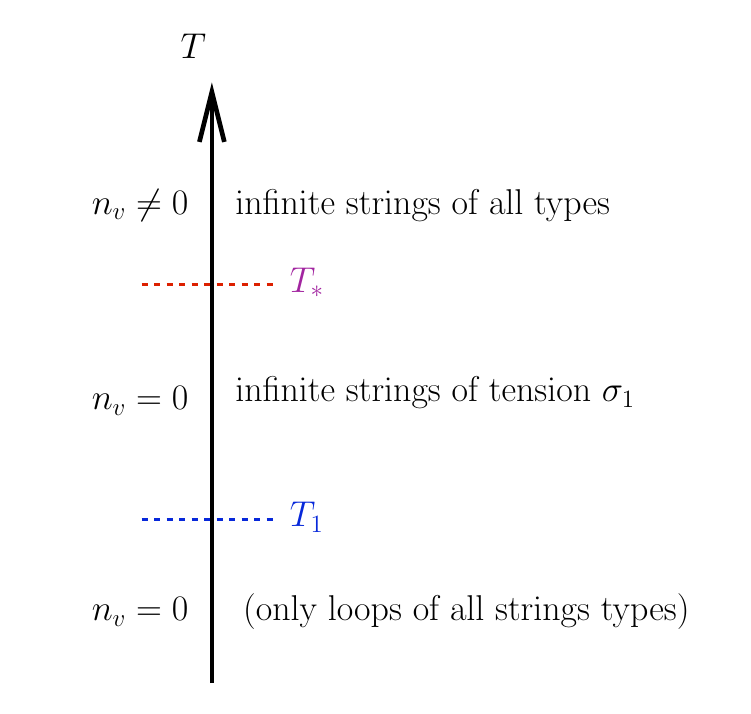}}
\caption{Different critical temperatures for our simplified model
of cosmic superstrings (with tensions $\sigma_1 < \sigma_2 <
\sigma_3$) with Y-junctions. The lower Hagedorn temperature $T_1$
is determined by $\sigma_1$ whereas the higher Hagedorn
temperature $T_*$ is a determined by all the $\sigma_{\alpha}$
$(\alpha=1,2,3)$. $n_v$ denotes the density of vertices (or
Y-junctions) joining infinite strings at temperature $T$.}
\label{fig:1}
\end{figure}

In subsequent sections we derive and analyse the phase structure
of a {\it three}-string model with junctions. This is a reduced
model of realistic cosmic superstrings, for which $(p,q)\in
\mathbb{Z} \times \mathbb{Z}$ form a doubly-infinite family. Since
string tension (or energy/unit length) increases with $p, q$, all
but low values will be suppressed at high temperature. We
therefore adopt the simplest non-trivial scheme, taking the two
lightest strings and their bound-state (and anti-strings), all
which have different tensions $\sigma_\alpha$, $\alpha=(1,2,3)$. For
example, depending on parameters, these could be the $(1,0)$,
$(0,1)$ and $(1,1)$ strings.
 We show that as the system is heated,
the lightest tension strings first undergo the Hagedorn
transition, despite the presence of Y-junctions. Conversely, at
low temperatures, only the lightest strings remain, before they
disappear into loops. Our results are summarized in figure
\ref{fig:1}.

These conclusions may have important consequences for $(p,q)$
string networks in that, if only the lightest strings remain after
a non-adiabatic quench, no significant r\^ole would be played by
the junctions whose properties have been studied so extensively.
The dynamics would then be that of a single string type with no
junctions (though there may be loops containing strings of
different types; as explained below, our analysis is limited to
infinite strings). This is not an idle proposition in that,
although our analysis in this paper assumes adiabatic behaviour,
we have learned elsewhere that universality classes of equilibrium
systems at their adiabatic transitions can become universality
classes of non-equilibrium systems at fast quenches \cite{Kibble}:
these points will be the content of a separate paper.  Other works \cite{us1,
Avgoustidis:2005nv,Jon,Tye:2005fn} based on studying the dynamics of
string networks with junctions also suggest that at late times only the lightest strings may remain.

The paper is set up as follows.  In section \ref{sec:underst} we
first review some relevant aspects of string statistical mechanics
in the simplest case: {\it one type} of string and {\it no} junctions.  In particular,
the duality between strings and fields is discussed.
In section \ref{sec:XY} we still consider only strings of a single tension and type, but
now these are allowed to meet at a junction.  This section paves the way for section \ref{sec:3}
in which we consider the general case of strings of three different tensions $\sigma_\alpha$ and types,
meeting at junctions.

As explained in section \ref{sec:XY} there is significant complexity involved in adding junctions
when discussing string statistical mechanics, and hence this section is central to the development of the paper.
Furthermore, technically, junctions can be introduced in different ways, and as a result we are forced to discuss in detail
two specific models (`bosonic' and `fermionic') to do so.
While bosonic models are closer to the physical system we eventually wish to describe (and discussed in
section \ref{sec:underst} when there are no junctions), only fermionic models can be generalized to the
three-string case of section \ref{sec:3}.  At the end of section \ref{sec:XY} we compare these two models,
and conclude that they both essentially agree in their phase structure.  This justifies the use of fermionic
models in section \ref{sec:3} where the analysis resulting in the conclusions
drawn in figure \ref{fig:1} is straightforward.  Finally, we also show, following ideas from
QCD, that the string system with junctions can be rewritten a generalised
spin model (XY model).

\section{Understanding the Hagedorn transition}
\label{sec:underst}

In this section we
discuss the nature of the Hagedorn
transition for strings of a {\it single} type, with tension $\sigma$, and {\it no} Y-junctions.

As mentioned in the introduction, we proceed by using the
duality between string configurations and fields to write the
partition function for the string network as that of an effective
field theory \cite{Stone}. As a result, the Hagedorn transition
can be mapped onto a transition of the effective field.
Furthermore, provided the right questions are asked, one can work
with the canonical rather than the microcanonical ensemble.

Consider a classical {\it static} picture of {\it
non-interacting} strings in $D$-spatial dimensions.  These are taken to lie on a
hypercubic lattice of spacing $a$, and the energy
$E$ of the strings only depends on the total string length $L$
through $E = \sigma L$. Near the critical temperature,
correlations are large and the details of the lattice structure
should be unimportant. We also assume that the network can be
thought of as a set of random walks.
Now recall
the {\it duality} between (non-oriented) Brownian paths in $D$ spatial dimensions and  a
scalar field $\varphi$ of mass $m$, as exemplified by the identity
 \ba
 &&\langle\varphi({\bf x})\varphi({\bf 0})\rangle =
  \nn
  \\
 && \int_0^{\infty} d\tau\,e^{-\tau m^2}\int^{{\bf x}({\tau})
 = {\bf x}}_{{\bf x}({0}) = {\bf 0}}{\cal D}{\bf x}\,\exp\bigg[-\int_0^{\tau}d\tau '\,
 \frac{1}{4}\bigg(\frac{d{\bf x}}{d\tau'}\bigg)^2\bigg].\nn
 \ea
This identity can be used to construct an effective {\it action} (or, more accurately, a free
energy) for the string partition function $Z$ at temperature $T =
\beta^{-1}$ in terms of $\varphi$ as \cite{Stone}
 \be
 Z = \int{\cal D}\varphi\,\exp\bigg[-\int dx^D\bigg(\frac{a^2}{4D}(\nabla\varphi)^2
 +\frac{1}{2}M^2\varphi^2\bigg)\bigg],
 \label{dualZ}
 \ee
where the mass term is
 \ba
 M^2  = \sigma a\beta \bigg(1 -
 \frac{T}{T_H}\bigg).\nn
 \ea
The Hagedorn transition temperature, $T_H = \beta_H^{-1}$, is the
solution to
 \be
 J(\beta)\equiv e^{-\beta\sigma a} = \frac{1}{2D}.\label{J}
 \ee
The normalisation of $\varphi$ has been chosen here so that $M^2$ is dimensionless.
(Note that one would have recovered the same temperature $T_H$ for a gas of strings by
counting single-loop configurations on the lattice \cite{Patel1,EdRay3}).

It is important to observe  that {\it below} the Hagedorn transition
$T<T_H$, $\varphi$ is a massive free field with $M^2$ {\it
positive}. For $T>T_H$, with $M^2<0$, it describes a {\it
tachyon}.  Here fluctuations are large and for this reason the
canonical ensemble often dropped in favour of the microcanonical
ensemble \cite{Mitchell1}. However, in the conventional picture of
spontaneous symmetry breaking we are familiar with the way in
which tachyons describe instabilities (in field space); they are
understood as corresponding to an inappropriate choice of ground
state, the true ground states appearing naturally once
back-reaction is taken into account.

For example, the inclusion of a repulsive point-interaction modifies the free energy to \cite{Stone}
 \be
 S = \int dx^D\bigg(\frac{a^2}{4D}(\nabla\varphi)^2
 +\frac{1}{2}M^2\varphi^2 +\lambda\varphi^4\bigg),
 \label{S0eff}
 \ee
thus permitting $\langle\varphi\rangle$ to remain finite for
$T>T_H$. For our $(p,q)$ networks, the system has more
complicated interactions than such a simple local repulsion. In
particular, were the strings allowed to interact at Y-junctions,
we would expect them to induce additional cubic $\mu\varphi^3$
terms
--- as we shall see in a different context below. However, the
general implications are much the same.

The vanishing of the order parameter $\langle\varphi\rangle$ at
$T\leq T_H$ can be understood in the following way. Examination of
the partition function shows that total string density is
proportional to $\langle\varphi^2\rangle$, whereas
$\langle\varphi\rangle^2$ measures the density in infinite string
(i.e.~string that crosses space) \cite{Stone,Ray}. It is the
vanishing of {\it infinite} string that characterises the Hagedorn
transition, and not the vanishing of string.

Although large loops are energetically unfavourable, some loops
will always exist below the transition (in an adiabatic limit).
Superficially, free energies like (\ref{S0eff}) look like those of
high-temperature quantum field theories on dimensional
compactification. Either from calculating the thermal propagator
for excitations at the relevant groundstate or by counting
 microstates of a loop gas we get  the same result that, in the
 vicinity of the transition, the loop distribution is dominated by
 the smallest possible loops (the ultraviolet limit) \cite{Ray}.

\section{Mean field transitions; XY models}
\label{sec:XY}

As discussed in section \ref{sec:underst}, we anticipate that Y-junctions will induce cubic
interaction terms in the dual field theory.  However, we do not know how to
introduce them in the exact framework of section
\ref{sec:underst}, even when the junctions are between strings of
the {\it same type and tension} $\sigma$ --- the setup considered in the
present section.

In this section we discuss a mean-field procedure
which allows junctions to be incorporated, and which shows how such
cubic interaction terms arise.  As in section \ref{sec:underst},
one can then construct an analogue effective potential, $V(\varphi
)$, for a field $\varphi$, whose vanishing describes the
transition.  Unfortunately, it is not possible to
extend this construction to the full effective action, and as a result
it is not possible to identify the field fluctuations that
describe finite loops: our analysis is restricted to infinite
string and the transitions triggered by its creation. Nonetheless,
knowing that loops are there enables us to complete the picture,
qualitatively.  It is the mean field procedure presented in this section which
will be generalised to the three-string model in section \ref{sec:3}.

Again we work in $D$ spatial dimensions, on a periodic hyper-cubic
lattice of $N$ sites and lattice size $a=1$. Let $i$ label a
lattice site, and $\mu=1,\ldots,D$ the (positive) unit vectors in
$D$-dimensional space.
There is now a technical complication,
related to how we
allow the strings to populate the lattice. Although there is an
energetic penalty in having more than one string on a link, in the
first instance we do not wish to restrict the number to unity. To
do so could imply an effective repulsion between strings that is a
lattice artefact, and which might induce misleading terms in the
effective potential for the analogue field $\varphi$. Without this
restriction the models are termed `bosonic'.

Models in which at most one string (of any type) can lie on a link
are termed `fermionic'. In practice we shall find, when we come to
mimicking $(p, q)$ strings, that only a fermionic model can
accommodate junctions of three string types.

An important result of this section is that our concern about
fermionic models is largely unjustified (though we feel it is necessary, for
reasons of clarity,  to discuss it in detail): both bosonic and
fermionic models essentially agree for the small $\varphi$
values that are relevant for transitions, and for which the mean
field approximation is more reliable.
Further, both of these models rewrites the string system as a
generalised XY model, permitting us to think of the Hagedorn
transition as one of spin ordering. This suggests ways of going
beyond the mean field approximation, although we shall not do so here.

\subsection{Bosonic models}

 With conventional lattice notation, let $n_{i,\mu}^{+}$ ($n_{i,\mu}^{-}$) be the number ($0,1,2,\ldots$) of
 strings
(anti-strings) on the link between the lattice points $i$ and
$i+\mu$.

For strings with no junctions, the Hamiltonian
 \be H = \sum_{i=1}^N
\sigma \sum_\mu (n_{i,\mu}^+ + n_{i,\mu}^-)
 \label{HB}
 \ee
 gives the requisite energy $E=\sigma L$ to a network of total
 length $L$.

Now, depending on the string network we wish to model, there is more than one way to proceed.
We discuss the mean field potenial in each case, making links with sections \ref{sec:underst} and \ref{sec:3}.

 \subsubsection{Massless junctions}

First we allow the strings to have $N_v$-fold {\it massless} junctions i.e.~no
 extra cost in energy. [We are primarily concerned with $N_v =
 3$.]
 Since the junctions considered are
massless they do not appear in the Hamiltonian, which is still
given by (\ref{HB}).

Rather, the existence of junctions imposes constraints on the $n_{i,\mu}^{+}$
($n_{i,\mu}^{-}$). Junctions or anti-junctions are permitted on
site $i$ provided the flux into that site is an integer multiple of $N_v$:
 \be
 \alpha_i \equiv \sum_\mu \left[ (n_{i,\mu}^+
- n_{i-\mu,\mu}^+) - (n_{i,\mu}^- - n_{i-\mu,\mu}^-) \right] = 0
 \; {\rm mod} \; N_v \label{conb},
 \ee
a constraint which can be implemented through
 \ba
 \delta_{\alpha=0 \; {\rm mod} \; N_v} = \frac{1}{N_v}
 \sum_{k_i=1}^{N_v} e^{i \alpha \theta_i} \qquad {\rm where} \qquad
 \theta_i = \frac{2\pi k_i}{N_v}.\nn
 \ea
 Using this representation in the canonical partition function
 \ba
 Z
= \sum_{n_{i,\mu}^\pm} e^{-\beta \sigma \sum_{i,\mu}(n_{i,\mu}^+ +
n_{i,\mu}^-)} \left(\prod_i \delta_{\alpha_i=0 \; {\rm mod} \; N_v}
\right) \nn
\ea
enables us to write $Z$ as
 \ba
 Z &=& \prod_i \frac{1}{N_v} \sum_{k_i} \left(
\sum_{n_{i,\mu}^+} e^{-\sum_{i,\mu}[\beta \sigma n_{i,\mu}^{+} + i
(\theta_{i+\mu}-\theta_{i})n_{i,\mu}^+]} \right)\times \nn
 \\
 &\times&\left(
\sum_{n_{i,\mu}^{-}} e^{-\sum_{i,\mu}[\beta \sigma n_{i,\mu}^{-} - i
(\theta_{i+\mu}-\theta_{i})n_{i,\mu}^-]} \right), \nn \ea
where the different signs in front of the lattice variables $\theta_{i}$
in the two terms in round brackets reflect the signs in (\ref{conb}).
The summations can be performed, to obtain
\ba Z =
 \left(\frac{1}{N_v}\right)^N  \sum_{k_i} e^{-\sum_{i,\mu}
\ln(1+J(\beta)^2-2J(\beta)
 \cos(\theta_i-\theta_{i-\mu}))}
 \nn
 \ea
where $J(\beta) = e^{-\beta \sigma}$ as in (\ref{J}). That is, the
Hamiltonian of the network is, up to a constant,
  \be
\beta H= \sum_{i,\mu}
\ln[1+ J(\beta)^2-2J(\beta)
 \cos(\theta_i-\theta_{i-\mu})].
 \ee

It is not possible to evaluate $Z$ exactly.
Hence we resort to the mean field approximation scheme (see for example
\cite{Kleinert}), which consists of introducing a trial
Hamiltonian $H_0$ in which each variable of the system is
decoupled from the other but
depends on an external constant source $\varphi$. An obvious
choice here is
\be
 H_0(\varphi) = - \frac{\varphi}{\beta} \sum_i \cos
 \theta_i \; .
 \ee
 On writing
 \ba
 H = H_0(\varphi) + [H - H_0(\varphi)],\nn
 \ea
 then
 \ba
 Z &=& \sum_{{\rm config}}
e^{-\beta H_0(\varphi)} e^{-\beta[H -H_0(\varphi)]}
\nn
\\
&=&
Z_0(\varphi)
\left\langle e^{-\beta[H -H_0(\varphi)]}\right\rangle_0 \nn
\\
&\geq& Z_0(\varphi) e^{-\beta \langle H -H_0(\varphi)\rangle_0
} , \nn
 \ea
 where the zero subscript denotes $\varphi$-dependent averaging with regard to $H_0(\varphi)$.
 As a result the free energy $F = -T\ln Z$ satisfies
 \be
F(\varphi) \leq NV(\varphi)\equiv F_0(\varphi) + \langle
H\rangle_0 -\langle H_0(\varphi)\rangle_0,
\label{interV}
 \ee
where $V(\varphi )$ is the mean field effective potential (and $F_0 = -T\ln Z_0$).
 Our aim is then to minimize $V$ in order to find $\varphi_{min}$, which
 determines the density of infinite string (see below).

We now carry out the calculation explicitly in the case of Y-junctions for which $N_v=3$.  Then
\ba Z_0(\varphi)  = \left[\frac{1}{3} \left(\sum_{k=1}^3
e^{\varphi\cos(2\pi k/3)} \right) \right]^N =
\tilde{I}_0(\varphi)^N \nn \ea
where
 \ba
 \tilde{I}_0
= \frac{1}{3}\left(e^\varphi + 2 e^{-\varphi/2}\right).
 \nn
 \ea
Now use the results that
 \be
 \langle \ln (1 +p^2 - 2 p \cos \theta) \rangle = -2
\sum_{m=1}^\infty
 \frac{p^m}{m}
 \langle \cos m\theta \rangle, \qquad (|p|<1)
 \label{lnBoson}
 \ee
for all measures, and that
 \be
 \langle\cos m\theta \rangle_0 = \frac{\tilde{I}_m (\varphi)}{\tilde{I}_0 (\varphi)}
 \ee
 for the case in point, where
 \ba
  \tilde{I}_m (\varphi)&=&
\frac{1}{3} \sum_k e^{\varphi\cos(2\pi k/3)} \cos( 2\pi m k/3)
 \nn
 \\
 & =&
\frac{1}{3} \left(e^\varphi + 2 e^{-\varphi/2}\cos(2\pi
m/3)\right)\nn
 \ea
is a discrete version of the Bessel function. Hence, using (\ref{interV}) we obtain
 \ba
 \beta V(\varphi) &=&  -\ln(\tilde{I}_0(\varphi)) +
 \varphi
\left(\frac{\tilde{I}_1(\varphi)} {\tilde{I}_0(\varphi)} \right)
\nn
\\
&& \qquad -2D \sum_{m=1}^\infty \frac{J(\beta)^m}{m}
\left(\frac{\tilde{I}_m(\varphi)} {\tilde{I}_0(\varphi)}
\right)^2. \label{Bvertices}
 \ea
 The periodicity (modulo 3) of the $\tilde{I}_m(\varphi)$ enables us to perform the
 summation explicitly, to give
 \ba
 \beta V(\varphi) &=&  -\ln(e^\varphi +
2e^{-\varphi/2}) + \varphi\left(\frac{e^\varphi - e^{-\varphi/2}}
{e^\varphi + 2e^{-\varphi/2}} \right)
 \nn
 \\
  && \qquad - 2DG(\beta) \left(\frac{e^\varphi -
e^{-\varphi/2}}{e^\varphi + 2e^{-\varphi/2}}\right)^2
 \label{vertex}
 \ea
where
 \ba
 G = \frac{1}{3} \ln \left(
\frac{1+J+J^2}{(1-J)^2} \right ) = J + \frac{1}{2}J^2 + .... \nn
 \ea
for small $J(\beta)$.

Notice that, because the sum over $m$ in (\ref{Bvertices}) just reproduces the first term
with a modified coefficient, $V(\varphi)$ of (\ref{vertex}) can be shown to be {\it
exactly} the mean-field potential arising from the Hamiltonian
 \be
 H^{disc}_{XY} = -\frac{G(\beta)}{\beta} \sum_{i,\mu}
{\bf{s}}_i \cdot {\bf{s}}_{i+\mu}. \label{HXY} \ee
i.e.~the Hamiltonian for a system of unit spins in the plane with
nearest neighbour interactions in which their relative angles are
constrained to multiples of $2\pi /N_v$ (here $N_v=3$); a discrete
XY model. The mean field trial Hamiltonian $H_0$ in this case is $
 H_0(\varphi) =  - \frac{\varphi}{\beta}{\bf n}.\sum_i{\bf{s}}_i$
 for an arbitrary unit vector {\bf n}
in which the spins are decoupled;  in other words, an external
magnetic field proportional to $\varphi$.

In order to understand the phase structure of the model (either as a spin system or as a gas of
strings with junctions),
consider first the series expansion of $V(\varphi)$;
 \be
\beta V(\varphi) = \frac{1}{2}m^2\varphi^2  +
\frac{1}{3}\mu\varphi^3 + \frac{1}{4}\lambda\varphi^4 + \ldots,
\label{quartic} \ee
 up to constant terms,
 where
 \be
  m^2 = \frac{1}{2}(1-2DG), \mu = \frac{1}{4}(1-3DG), \lambda = -\frac{3}{16}(1-2DG).\label{Bstrings} \ee
  Observe that the field becomes massless at the temperature for which $2DG(\beta) =1$, which is in
  good
agreement with the Hagedorn temperature of the free dual theory
 of (\ref{dualZ}) since $G(\beta) \simeq J(\beta)$ for
 $J = 1/2D\ll 1$.   Furthermore, as anticipated, the Y-junctions have
 induced a cubic term in the potential. In addition they have also induced a quartic interaction,
 vanishing when the field
 becomes massless, that is repulsive when the field becomes tachyonic.

As a result of the cubic term, the potential in equation (\ref{vertex}) can be
shown to have  a weak first order phase transition. The critical
temperature, however, cannot be obtained from (\ref{Bstrings}) as
it occurs at values of $\varphi\simeq 1$. Numerically, however,
one finds that $2G_{crit}(D=3)\simeq 0.31$ and $2G_{crit}(D=4)\simeq
0.23$. We shall not consider the first order transition further,
since it is not reliably robust against rapid quenches which is
what we ultimately have in mind.

\subsubsection{Massive junctions}

Alternatively one might want to model string networks with massive junctions --- that is,
is to introduce junctions with an energy cost $v$. (These can model massive monopoles, which may be formed
at the vertex in different symmetry breaking schemes \cite{VV}.)  We can then recover massless vertices by taking $v\rightarrow 0$.
Furthermore this construction allows one to calculate the average
density of vertices at temperature $T$, by simply differentiating
$Z$ with respect to $v$. This will be discussed in section
\ref{sec:3}.

To add massive vertices, we allocate a vertex number $p_i^{\pm} =
(0,1,2\ldots)$ to each lattice site, constrained by
 \ba \alpha_i &
 \equiv& \sum_\mu \left[ (n_{i,\mu}^+ - n_{i-\mu,\mu}^+) -
 (n_{i,\mu}^- - n_{i-\mu,\mu}^-) \right] \nn
\\
 && \qquad \;+ \; 3 (p_i^+ -
 p_i^-) = 0
 \label{heavyVB}
 \ea
for Y-junctions,
while the Hamiltonian acquires an extra term
 \be
 H_I = \sum_{i=1}^N  v (p_i^+ + p_i^-).
 \ee
  Performing the sums over the $n^{\pm}_{i,\mu}$ and the $p_i^{\pm}$ leads to a
 Hamiltonian
 \ba
 \beta H &=& -\sum_{i,\mu}\ln [ 1 + J^2(\beta) - 2J(\beta)
\cos(\theta_{i+\mu} - \theta_{i})] \nn
 \\
  &-& \sum_i\ln [ 1 + K^2(\beta)
- 2K(\beta) \cos 3\theta_{i}], \label{HB2}
 \ea
 where the $\theta_i$ are now {\it continuous variables}, the Lagrange multipliers that arise from
 imposing the constraints
 \be
 \delta_{\alpha_i,0} = \frac{1}{2\pi} \int_0^{2\pi} d\theta_i
 e^{i \alpha_i \theta_i}.
 \ee
Also we have defined
\be {K(\beta)}=e^{-\beta v}, \label{Kdef}
\ee
analogously to $J$
in (\ref{J}).
Then, carrying out the same mean field treatment as above yields
\ba
 \beta V^{(K)}(\varphi) &=&  -\ln({I}_0(\varphi)) + \varphi\left(\frac{{I}_1(\varphi)} {{I}_0(\varphi)}
 \right)
 \nn
 \\
 && -2D \sum_{m=1}^\infty \frac{J(\beta)^m}{m}
 \left(\frac{{I}_m(\varphi)} {{I}_0(\varphi)} \right)^2
 \nn
 \\
 &&
 - 2 \sum_{m=1}^\infty \frac{{K(\beta)}^m}{m} \left(\frac{{I}_{3m}(\varphi)}
{{I}_0(\varphi)} \right),
 \label{V0boson}
 \ea
where the $I_m$ are (continuous) Bessel functions.

For non-zero $K$ cubic terms arise from the $I_{3}$ Bessel
function, to give rise to a potential of the form (\ref{quartic}),
with coefficients
\be
m^2 = \frac{1}{2}(1-2DJ), \; \; \;  \mu = -\frac{K}{8}, \; \; \; \lambda = -\frac{3}{16}\bigg(1-\frac{8DJ}{3}\bigg).\label{Bstrings2} \ee
As expected, we have tachyonic instability at $J= 1/2D$ and a
cubic term in the potential.

The slightly different behaviour of (\ref{Bstrings2}) and
(\ref{Bstrings}) is to be expected, since since we are
implementing the boundary conditions that count vertices
differently in the two cases: in other words, they correspond to different
implementations of the mean field approach. However, since the
mean field result is, strictly, an upper bound, we could, if we
wished,  only retain that solution that is numerically lower. In
practice, this is not necessary since there is close numerical
agreement at relevant temperatures. Massless junctions correspond
to taking $K=1$ for which $\mu = -1/8$, the value arising in
(\ref{Bstrings}) when $2DG = 1$. Further, a numerical study of
(\ref{V0boson}) shows that the transition tends to become first
order as ${K} \rightarrow 1$, in agreement with the discussion of
(\ref{Bstrings}).

\subsubsection{No junctions}

For continuous `bosonic' string with no junctions both approaches
give the identical result. In the first case, we eliminate
junctions by taking $N_v\rightarrow\infty$, whereby the discrete
Bessel functions are replaced by their continuous counterparts. In
the second, taking $v\rightarrow \infty$ ($K=0$) just recreates
the same series.

In each case, on expanding $V^{(0)}(\varphi)$ for small $\varphi$
we recover the second order transition at the Hagedorn temperature
$T_H$ of Section II (see equation (\ref{J})) when $2DJ(\beta) =
1$, and when the $\varphi$ field becomes tachyonic. However, it
can be seen that $V^{(0)}$ of (\ref{V0boson}) becomes unbounded
below as $T\rightarrow\infty$. This is not quite the behaviour of
(\ref{dualZ}), for which the potential is unbounded below for all
$T> T_H$, showing the limitations of the mean field approach for
very large $|\varphi|$. Nonetheless, this simple example shows how
the introduction of vertices induces interaction terms in the
effective potential to stabilise the ground states.

\subsection{Fermionic models}

We now consider the most simple `fermionic' models. It is these
which can straightforwardly be extended to the general three
string-type model of section \ref{sec:3}.  We will also address the concern raised at the beginning of this section: that
the `fermionic' model might add an effective repulsion between
strings, which could induce misleading terms in the effective
potential.  We will show that this is not the case.

Thus, we now
restrict the number of strings on each link to $n_{i,\mu}\in
\{0,\pm 1\}$. That is, the link from site $i$
to $i+\mu$ contains either a single string, a single anti-string,
or no string at all.

\subsubsection{No junctions}

With no junctions, the Hamiltonian is
 \be
H =\sum_{i=1}^{N}\sum_{\mu=1}^D \sigma  n_{i,\mu}^2, \label{Hg}
 \ee
 subject to the constraint
 \be
\alpha_i \equiv  \sum_{\mu} \left[ n_{i,\mu}-n_{i-\mu,\mu}\right]
=0.
 \ee
 Performing the sums over the $n_{i,\mu}$ leads to a
 Hamiltonian
  \ba
 \beta H = -\sum_{i,\mu}\ln [ 1 + 2J(\beta)
\cos(\theta_{i+\mu} - \theta_{i})],
\label{HF}
 \ea
 where the $\theta_i$ are again the Lagrange multipliers that arise from
 imposing the constraints
 \be
 \delta_{\alpha_i,0} = \frac{1}{2\pi} \int_0^{2\pi} d\theta_i
 e^{i \alpha_i \theta_i}.
 \ee

 Defining ${\bar J}$ by
 \be
 J =\frac{\bar J}{1 + {\bar J}^2},
 \label{barJ}
 \ee
 whereby $J(\beta)\approx {\bar J}(\beta)$ when $J\ll 1$, a similar calculation to that above
 (see also section \ref{sec:3}) shows that the mean-field potential
 is, for ${\bar J}<1$,
 \ba
&& \beta V^{(0)}_{F}(\varphi) = -  \ln I_0(\varphi) \nn
 \\
&& \qquad + \; \varphi
\left(\frac{I_1(\varphi)}{I_0(\varphi)}\right) + 2D
\sum_{m=1}^{\infty}
 \frac{(-\bar{J}(\beta))^m }{m}
\left(\frac{I_m(\varphi)}{I_0(\varphi)}\right)^2. \qquad
 \label{V0fermion}
 \ea
 The $\varphi$ field now becomes massless at
 $2D{\bar J}(\beta) = 1$, with a second order transition. With $J\approx{\bar J}$ this is slightly displaced from
 that of the bosonic strings but, at the qualitative level at which we are working,
 can be said to agree. Note that both
 potentials (\ref{V0fermion}) and (\ref{V0boson}) show a $\mathbb Z_2$
 symmetry under $\varphi\rightarrow -\varphi$ that is broken {\it
 above} $T_H$, and restored {\it below} $T_H$, contrary to the
 usual pattern of symmetry breaking, but as in section \ref{sec:underst}.

 On comparing (\ref{V0fermion}) with (\ref{V0boson})  we see that
 they differ in that the former has alternating signs in
 the Bessel function expansion, whereas the latter does not.
Because higher terms in the series in powers of ${\bar J}(\beta)$
become significant only at increasingly large $\varphi$, the
artificial repulsion induced by the `fermionic' assumption (that is, of no
more than one string per link) is a large-$\varphi$ effect in the
mean field approximation, and hence where the approximation is at its least
reliable.
However, since the transitions are determined by small $\varphi$,
we can use either.  This is an important result of this section.

In fact, for $J$ small, both approximate the mean-field  potential
of the XY-model, with spin-spin Hamiltonian
 \ba
 H_{XY} &=& - \frac{1}{\beta}\sum_{i,\mu}
2J\cos(\theta_{i+\mu} - \theta_{i})
 \nn
 \\
 &=& -\frac{2J}{\beta} \sum_{i,\mu}
{\bf{s}}_i \cdot {\bf{s}}_{i+\mu}.
 \ea
 This follows from expanding (\ref{HF}), for which
 \be
V_{XY}(\varphi) =  - \ln I_0(\varphi) + \varphi
\left(\frac{I_1(\varphi)}{I_0(\varphi)}\right) - 2DJ(\beta)
\left(\frac{I_1(\varphi)}{I_0(\varphi)}\right)^2,
 \label{VXY} \ee
showing a second order transition at $2DJ(\beta) = 1$. Rather than
just perform a series expansion in $\varphi$ as in
(\ref{Bstrings}), more generally we see that extrema of
$V_{XY}(\varphi)$ satisfy
 \be
 {\bar\varphi} - 4DJ(\beta)u({\bar\varphi}) = 0,
 \label{cubic}
 \ee
 where $u(\varphi) = I_1(\varphi)/I_0(\varphi)$.

 $\bar\varphi = 0$ is
 always a solution to (\ref{cubic}). For $2DJ(\beta) > 1$ there is a further pair of solutions,
 $\pm{\bar\varphi},\,\,\,{\bar\varphi} >0$, which are the minima.
 We note, for future use, when we need to count extrema, that (\ref{cubic}) behaves like the
 cubic equation obtained from just retaining terms up to
 $O(\varphi^4)$ in the expansion of the potential in the existence
 of three roots.  The inclusion of higher terms in the series in $\bar{J}$ does
not seem to affect this empirically and it is not necessary to go
beyond the XY model, now and hereafter.

When the XY model is a good approximation we could, in principle,
use known results about it without resorting to the mean-field
approximation. In practice, we know of no work on the generalised
XY models appropriate to the three-string models (with or without junctions) and stay with the
mean-field approximation.

To give a meaning to ${\bar\varphi}$ we note that the average
density of (infinite) strings is proportional to
${\bar\varphi}^2$, as anticipated, given by
 \be
\rho = \frac{1}{N} \langle \sum_\mu n_{i,\mu}^2 \rangle
 = -J(\beta)\beta
\frac{\partial V_{XY}}{\partial J} =
\frac{{\bar\varphi}^2}{4DJ(\beta)}.
 \label{rhostrings}
 \ee

\subsubsection{Massive junctions}

 We end this section by
including  Y-junctions in the fermionic model (still of a single
string type).  Given that the occupation numbers are limited to
$0,\pm1$, there is no analogue of the mod 3 description for
massless vertices discussed in the bosonic case (see equation
(\ref{conb})).  We therefore consider massive vertices.  There is now a single vertex
number $p_i=\{0,\pm 1\}$ constrained by
  \be
\alpha_i \equiv  \sum_{\mu} \left[ n_{i,\mu}-n_{i-\mu,\mu}\right]
+ 3p_i =0
 \label{con}
 \ee
with the Hamiltonian acquiring an additional term
 \be
  H_I = \sum_i v p_i^2.
  \ee
On defining ${\bar K}$ by
 \be
K =\frac{\bar K}{1 + {\bar K}^2},
 \label{barK}
 \ee
the mean-field potential
 is, for $({\bar J},{\bar K}<1)$,
 \be
 \beta V^{(K)}_{F}(\varphi) =\beta V^{(0)}_{F}(\varphi) + 2
\sum_{m=1}^{\infty}
 \frac{(-\bar{K}(\beta))^m }{m}
\left(\frac{I_{3m}(\varphi)}{I_0(\varphi)}\right)^2. \qquad
 \label{Vfermion}
 \ee
 [This follows from the generalisation of
 (\ref{lnBoson}), used earlier in (\ref{V0fermion}) that, up to a constant,
 \ba
 &&\langle \ln (1 + 2K \cos \alpha) \rangle =
  -2 \sum_{m=1}^\infty\frac{ \left( - \bar{K} \right)^m}{m}
 \langle \cos m\alpha \rangle \qquad
 \label{horrible}
 \ea
 for all measures and $\bar{K}<1$, together with the specific result
 \ba
\langle \cos m\theta \rangle_0 \equiv \frac{\int
\frac{d\theta}{2\pi} e^{ \varphi \cos\theta } \cos m\theta}{\int
\frac{d\theta}{2\pi} e^{ \varphi \cos\theta }} &=& \frac{I_m(\varphi
)}{I_0(\varphi )}
\nn
\ea
 for our choice of $H_0$.]
We note that unfortunately, for a simple cubic lattice, the
requirement that $\bar{K}<1$, necessary for convergence of the
series in (\ref{horrible}),
 imposes $K<1/2$. Hence that the mean field approximation is not valid for light vertices in the fermionic case
 (as opposed to the bosonic one in (\ref{V0boson})).  We consider this constraint to be an artefact of the lattice
 fermionic approximation.

Despite that, note that mean field potential (\ref{Vfermion}) leads to an XY model in the presence of an external
 source
 \cite{Patel1,MR} in which we retain only the first term in the power series in ${\bar K}$ in (\ref{Vfermion})
 (or the first term in the series in $K$ in (\ref{V0boson})).
 As a result, there is always a second-order transition, as in the bosonic
 case.

Finally we also note that the density of string (\ref{rhostrings}) is unchanged by
 the inclusion of junctions.

\subsection{Summary of section \ref{sec:XY}}

In summary, in this section we have seen how the inclusion of Y-junctions in a
model of a single string type can provide the back-reaction
necessary to prevent tachyonic instability at the Hagedorn
temperature. Further, provided we restrict ourselves just to
infinite string, whose density is the order parameter, we can go
beyond the Hagedorn temperature, still with the canonical
ensemble.

We have also discussed two models, `bosonic' and `fermionic', and
shown that the concern raised about fermionic models at the
beginning of this section is unjustified: both models essentially
agree for the small $\varphi$ values that are relevant for
transitions, and for which the mean field approximation is more
reliable.

We have also shown how the value $\bar{\varphi}$ of the field at
the minimum of the effective potential is related to the density
of infinite strings in the system. As we discuss in section \ref{sec:3}, it is
equally apparent that the density of vertices is also determined
by $\bar{\varphi}$ and obtained by differentiating the partition
function with respect to $v$.

With this behind us, we now consider the case of three different
string types with Y-junctions, as a model for $(p,q)$ strings. We
note that, oddly, the analysis of QCD confinement of
\cite{Patel1,Patel2,MR}, that we have called upon in this paper,
was performed in the context of a single-string model, not
permitting `colour'. Although this was not our intention, a more
realistic description of QCD is given by the model that follows,
in the limit of equal tensions, in which our three string types
correspond to coloured flux tubes.

\section{Three strings, fermionic model}
\label{sec:3}

The basics of our model are the following.

As stated in the introduction, we model the $(p,q)$ string network
by a network of three different types of fundamental strings,
labelled by $\alpha = 1, 2, 3$ as red, green and blue, say.
Generally the strings also have different tensions
$\sigma_\alpha$.  The strings do not interact with each other (nor
with themselves), except at a Y-junction (or vertex) which is
defined to be a point at which three strings of {\it different} colours
meet.

Following equation (\ref{quartic}), our expectation is that the effective
potential will take the generic form
 \ba
 \beta V(\varphi_1,\varphi_2,\varphi_3) &=& \sum_{\alpha}\bigg[
 \frac{1}{2}m_{\alpha}^2\varphi_{\alpha}^2 +
 \frac{1}{4}\lambda_{\alpha}\varphi_{\alpha}^4
 \bigg]
 \nn
 \\
 && \qquad + \, \mu \, \varphi_1\varphi_2\varphi_3 + ...\label{V3}
 \ea
 Potentials of the type (\ref{V3}), with temperature-dependent
 coefficients,
 have been studied in other contexts e.g.~transformations of vortex types in superfluid $^3$He \cite{Volovik}.

 We know that (\ref{V3}) is valid if Y-junctions are excluded, when $\mu = 0$.
 In this case, from the single string models
\be
m_{\alpha}^2 \propto (1-2DJ_{\alpha}(\beta )),
\ee
with $J_\alpha = e^{-\sigma_{\alpha}\beta}$.
In the following discussion we suppose that
\be
\sigma_1 \leq \sigma_2 \leq \sigma_3 \qquad \Longleftrightarrow \qquad
J_1 \geq J_2 \geq J_3.
\ee
The critical ${J}^{crit}_\alpha = 1/(2D)$ define three critical
inverse temperatures $\beta_\alpha = T_{\alpha}^{-1}$ with
 \be
 \beta_3 < \beta_2 <\beta_1
 \ee
 in the vicinity of which $m_{\alpha}^2\propto (1-T/T_{\alpha})$.
 That is, with no interactions we expect three
 sequential Hagedorn transitions as, on cooling, the heavier
 strings disappear from the picture, leaving the lightest until
 last before it disappears in turn, leaving just small loops.

 Our aim is to understand the effect that Y-junctions have on this
 picture.

In practice, we are not able to recreate (\ref{V3}) in a bosonic
model with coloured Y-junctions, with arbitrary numbers of strings
on each link.  (The reason is that we are unable to write down a
generalised form of the constraint (\ref{conb}) in the 3-string case.)  We therefore restrict ourselves
to a fermionic model, in which there is at most one string of each
type on a link. As discussed in the previous section, we
expect that the effective repulsion this implies can be ignored at
small field values. As in the case of the single string type, in
order to be able to use mean field theory we are obliged to give
the vertex a non-zero mass $v$.

 As before, we assume that the energy of
the different strings is proportional to their length $L$
($E_\alpha=\sigma_\alpha L$).  The different strings are described
respectively by the variables $n^{\alpha}_{i,\mu}$, which all take values in $\{ 0,\pm 1\}$. There are also
vertices, described by the variable $p_i \in \{ 0,\pm 1\}$,
joining strings of 3 different types.
The Hamiltonian of the system takes the same form as the for the
single string case,
 \be H = \sum_i \left[ \sum_\mu
\sum_{\alpha}\sigma_{\alpha}(n^{\alpha}_{i,\mu})^2 + v p_i^2
\right] \ee

We now
need to impose the constraint that a junction is where three different colour
strings meet: this is done by
 \ba
 \gamma^{\alpha}_i =\sum_\mu (n^{\alpha}_{i,\mu} - n^{\alpha}_{i-\mu,\mu}) + p_i &=& 0, \qquad \forall
 \alpha.
 \label{con2}
 \ea
 Although summing over $\alpha$ would essentially recreate the
 constraints (\ref{con}), equation (\ref{con2}) is more specific.
 In particular, (\ref{con2}) does not forbid different string types from lying
on top of each other.

As in the previous section, the constraints are imposed in the standard way through Lagrange
multipliers, which is equivalent to writing the Kroneker delta as
 \be
 \delta_{\gamma_i^{\alpha},0} = \frac{1}{2\pi} \int_0^{2\pi} d\theta^{\alpha}_i
 e^{i \gamma^{\alpha}_i \theta^{\alpha}_i}
 \ee
 (no $\alpha$ summation) for each $\gamma^{\alpha}$.
  Hence the partition function is
  \ba
  &&Z(\beta,v,\sigma_{\alpha}) =
  \nn
  \\
   &&= \int \prod_{i,\alpha} \frac{d\theta^{\alpha}_i}{2\pi}
\sum_{n_{i,\mu}} e^{-\sum_{i,\mu} \sum_{\alpha}[\beta
\sigma_{\alpha} (n^{\alpha}_{i,\mu})^2 + i
n^{\alpha}_{i,\mu}(\theta^{\alpha}_{i+\mu} -
\theta^{\alpha}_{i})]} \nn
\\
&& \qquad \times \sum_{p_i}e^{-\sum_{i}[\beta v p_i^2 + i
p_i\sum_{\alpha} \theta^{\alpha}_i]}
 \ea
  which, on carrying out the summations gives
  \ba
 Z(\beta,v,\sigma_\alpha) &=& \int \prod_{i,\alpha} \frac{d\theta^{\alpha}_i}{2\pi}
  \prod_{\mu} [ 1 + 2J_\alpha
\cos(\theta^{\alpha}_{i+\mu} - \theta^{\alpha}_{i})] \nn
\\
&& \qquad \times  [ 1 + 2K \cos(\sum_{\alpha}\theta^{\alpha}_i)]
\label{full} .
\ea
This corresponds to the Hamiltonian
 \ba
 \beta H = -\sum_{i,\mu,\alpha}\ln [ 1 + 2J_\alpha
\cos(\theta^{\alpha}_{i+\mu} - \theta^{\alpha}_{i})]
\nn
\\
-\sum_i\ln [ 1 + 2K \cos(\sum_{\alpha}\theta^{\alpha}_i)] \nn
\\
\approx \sum_{i,\mu,\alpha} 2J_\alpha \cos(\theta^{\alpha}_{i+\mu}
- \theta^{\alpha}_{i}) + \sum_i 2K
\cos(\sum_{\alpha}\theta^{\alpha}_i)\label{H3}
 \ea
 for small $J_{\alpha}$ and $K$.

 The mean field
treatment therefore contains three variational parameters
$\varphi_\alpha$.  Following the same steps as in section \ref{sec:XY},
the trial partition functions which decouple
different lattice sites are
\be Z_0^\alpha(\beta,\sigma_\alpha,\varphi_\alpha) = \int
\prod_{i} \frac{d\theta^\alpha_i}{2\pi} e^{\sum_i \varphi_\alpha
\cos\theta^\alpha_i } = \left[I_0(\varphi_\alpha)\right]^N,
 \ee
while the mean field effective potential is
 \ba
&&  \beta V(\varphi_{\alpha}) =\sum_{\alpha}\bigg[ - \ln
I_0(\varphi_{\alpha}) +
%\right.
  \nn
  \\
 && + \left. \varphi_{\alpha}
\left(\frac{I_{\alpha}(\varphi_{\alpha})}{I_0(\varphi_{\alpha})}\right)
+ 2D \sum_{m=1}^{\infty}
 \frac{(-\bar{J_{\alpha}})^m }{m}
\left(\frac{I_m(\varphi_{\alpha})}{I_0(\varphi_{\alpha})}\right)^2
\right] \nn
\\
&& + \;
  2 \sum_{m=1}^{\infty}
  \frac{(-\bar{K})^m}{m}
\left(\frac{I_m(\varphi_1)}{I_0(\varphi_1)}
\frac{I_m(\varphi_2)}{I_0(\varphi_2)}
\frac{I_m(\varphi_3)}{I_0(\varphi_3)}\right),
 \label{general} \ea
 where each
 $\bar{J}_{\alpha}$ is defined as in (\ref{barJ}), and
  ${\bar K}$ is given in (\ref{barK}).

As discussed in section \ref{sec:XY}, it is sufficient for our purposes to approximate $ \beta
V(\varphi_{\alpha})$ by the first term in the series of
(\ref{general}),
 \ba
&&  \beta V_{XY}(\varphi_{\alpha}) =\sum_{\alpha}\bigg[ - \ln
I_0(\varphi_{\alpha}) +
  \nn
  \\
 && + \left. \varphi_{\alpha}
\left(\frac{I_{\alpha}(\varphi_{\alpha})}{I_0(\varphi_{\alpha})}\right)
-2D J
\left(\frac{I_1(\varphi_{\alpha})}{I_0(\varphi_{\alpha})}\right)^2
\right] \nn
\\
&& - \;
 2K
\left(\frac{I_1(\varphi_1)}{I_0(\varphi_1)}
\frac{I_1(\varphi_2)}{I_0(\varphi_2)}
\frac{I_1(\varphi_3)}{I_0(\varphi_3)}\right).
 \label{3XY} \ea
This corresponds to making the small $J,K$ approximation in
(\ref{H3}). That is, the model (\ref{full}) is a generalised XY
model, consisting of three spin-like variables defined on each
lattice site $i$, making angles $\theta^{\alpha}_i$ with respect
to some fixed axis, interacting amongst themselves through the
$K$-dependent term.

We have achieved our goal in that, if we expand
$V_{XY}(\varphi_{\alpha})$ of (\ref{3XY}) (or, indeed the full
$V(\varphi_{\alpha})$ of (\ref{general})) in powers of
$\varphi_{\alpha}$ we recover the generic potential (\ref{V3}) as
the first few terms in the series.

However,  we can say more. As in our earlier examples, attaching a
nominal energy to each vertex allows us to calculate the density
of vertices. Specifically, the density of vertices on infinite
strings is
 \ba
n_v &=& \frac{1}{N} \langle \sum_i p_i^2 \rangle
 = -K\beta
\frac{\partial V_{XY}}{\partial K} \nn
\\
&\propto& \left(\frac{I_1({\bar\varphi_1)}}{I_0({\bar\varphi_1)}}
\frac{I_1({\bar\varphi_2)}}{I_0({\bar\varphi_2)}}
\frac{I_1({\bar\varphi_3)}}{I_0({\bar\varphi_3)}}\right)\propto
{\bar\varphi_1}{\bar\varphi_2}{\bar\varphi_3}
 \label{nv}
 \ea
 at the minimum
 $({\bar\varphi_1},{\bar\varphi_2},{\bar\varphi_3)}$ of $V_{XY}(\varphi_\alpha)$.
The small loops corresponding to the field fluctuations that are
invisible to our mean field analysis contain vertices not counted
in (\ref{nv}).

As in section \ref{sec:XY}, we now look for the extrema of the
potential in order to determine the density of infinite string and the density of vertices. As
expected from section \ref{sec:XY}, a full numerical analysis
(that we have performed) without the XY approximation does not
alter our qualitative conclusions and barely changes our
quantitative results.

As we noted earlier, in the main works on QCD
(\cite{Patel1,Patel2}) all flux strings were taken to be of a
single kind, leading to a very different potential, in which
$I_1(\varphi_1) I_1(\varphi_2) I_1(\varphi_3)$ is replaced by
$I_{3}(\varphi)$ for example. In particular, as we shall see later
for (\ref{general}), with equal tensions there is no first-order
transition when there are three string types.

\subsection{$K=0$: no vertices and three independent spins}

We have already anticipated the results for this simple case, but
it is helpful to see them in greater detail. For $K = 0$ the XY
model reduces to three independent, uncoupled, XY models with
$\mathbb{Z}_2 \times \mathbb{Z}_2 \times \mathbb{Z}_2$ symmetry
under $\varphi_{\alpha}\rightarrow - \varphi_{\alpha}$.
The extremal points are when
 \be
 \frac{\partial V_{MF}}{\partial \bar\varphi_\alpha}=0 \qquad
 \Leftrightarrow \qquad \frac{\bar\varphi_\alpha}{4DJ_\alpha} -
 u(\bar\varphi_\alpha)=0
 \ee
 where $u(\varphi) = I_1(\varphi)/I_0(\varphi)$ as before.  One
possible solution is always $\bar\varphi_\alpha=0$, the only real
solution if $2DJ_\alpha(\beta) < 1$.

If $2DJ_\alpha(\beta)
> 1$ then there are two further real solutions, denoted $\pm{\bar
\varphi}_\alpha$, where we take ${\bar \varphi}_\alpha >0$.  The
$3^3 = 27$ possible extrema $\varphi = (\bar\varphi_1,\bar
\varphi_2, \bar\varphi_3 )$ then break down into a non-degenerate
$\varphi = (0, 0, 0)$, three doubly degenerate solutions,
exemplified by $\varphi = (\pm{\bar\varphi}_1, 0, 0)$, three
fourfold degenerate solutions, exemplified by
$(\pm{\bar\varphi}_1,\pm{\bar\varphi}_2,0)$ and an eightfold
degenerate solution
$(\pm\bar{\varphi}_1,\pm\bar{\varphi}_2,\pm\bar{\varphi}_3)$. It
is sufficient to restrict ourselves to the positive sector
$\varphi_{\alpha}\geq 0$.

To determine which of these are maxima, which minima, and which
saddle points we need to calculate the eigenvalues of the Hessian
$ M_{\gamma \delta} = {\partial^2 V_{XY}}/{\partial \varphi_\gamma
\partial \varphi_\delta} $ at the extrema. An extremum is a
minimum if all are positive, and a maximum if all are negative.
Otherwise one is dealing with saddle points.

With $K=0$, the only non-zero entries are on the diagonal with (no
summation)
 \be
 M_{\alpha \alpha} = u'(\bar\varphi_\alpha)[1-4DJ_\alpha (\beta)
 u'(\bar\varphi_\alpha)] \label{second}.
 \ee
 For the case in hand the answer is very simple and very obvious.

\begin{enumerate}
\item $\beta > \beta_1 (> \beta_2,\beta_3)$.  In this range the
global minimum occurs at $\vec{\varphi} = (0,0,0)$.
\item $\beta_2 < \beta < \beta_1$. Now $(\bar{\varphi}_1,0,0)$ is
the {\it global minimum}.  [$(0,0,0)$ is a now saddle point.]

\item $\beta_3 < \beta < \beta_2$.  In this range it is easy to
see that $({\bar\varphi}_1,{\bar\varphi}_2,0)$ is the global
minimum.

\item $\beta < \beta_3$.  Here it is equally straightforward to
see that $(\bar{\varphi}_1,\bar{\varphi}_2,\bar{\varphi}_3)$ is
the global minimum, $(0,0,0)$ is a maximum, and all other points
are saddle points.

\end{enumerate}

As expected, as the temperature is increased infinite strings of
the lightest tension first are nucleated at $\beta=\beta_1$; then
those of the next lightest tension at $\beta=\beta_2$; and finally
the heaviest strings when $\beta=\beta_3$.  When one decreases the
temperature from a very high one, the opposite happens.

\subsection{$K \neq 0$: vertices and three coupled spins}

Let us now consider the effect of Y-junctions in the generalised
XY model of (\ref{3XY}).
For unequal $\sigma_{\alpha}$ the symmetry of $V_{XY}$ is now
explicitly broken from $\mathbb{Z}_2 \times \mathbb{Z}_2 \times
\mathbb{Z}_2$ to $D_2= \mathbb{Z}_2 \times \mathbb{Z}_2$,
generated by
 \ba
 P_1:\qquad \varphi_1\rightarrow\varphi_1, \qquad\varphi_2\rightarrow -\varphi_2,
 \qquad\varphi_3\rightarrow -\varphi_3\nn\\
 P_2:\qquad \varphi_1\rightarrow -\varphi_1, \qquad\varphi_2\rightarrow\varphi_2,
 \qquad\varphi_3\rightarrow -\varphi_3\nn\\
 P_3:\qquad \varphi_1\rightarrow -\varphi_1, \qquad\varphi_2\rightarrow -\varphi_2,
 \qquad\varphi_3\rightarrow\varphi_3\nn
 \ea
 If any tensions are equal the symmetry is correspondingly
 increased.
Imposing $\partial V_{XY}/\partial \varphi_\alpha=0$ gives (no
summation)
 \ba
 u'(\bar\varphi_\alpha) \left[ {\bar\varphi_\alpha} -
{4DJ_\alpha}(\beta) u(\bar\varphi_ \alpha)-2K(\beta)
u(\varphi_\beta)u(\bar\varphi_\gamma)\right] &=& 0
\nn
\\
\label{11}
 \ea
 where $\beta = (\alpha+1) \, {\rm mod} \, 3$, $\gamma =
(\alpha+2) \, {\rm mod} \, 3$. There are obvious solutions to these
coupled equations: $(0,0,0)$ for all $\beta$; $(\varphi_1,0,0)$
with $\varphi_1=\bar{\varphi}_1$ (the standard solution provided
$2DJ_1 >1$). The important point though is that {\it it is not
possible to have a solution with only, say $\bar\varphi_1=0$, and
the other two non-zero}.  One can see this from (\ref{11}), where
setting $\bar\varphi_1=0$ would require that one of the other two
$\varphi$'s must vanish.

At the extrema the Hessian has the same diagonal elements as in
(\ref{second}), but off-diagonal elements 
\ba
 M_{\alpha \beta} &=& -2K(\beta) u'(\bar\varphi_\alpha)
u'(\bar\varphi_\beta) u(\bar\varphi_\gamma)
 \ea
 We now evaluate these at
the different extrema identified above and discuss the consequences.
\\
\\
\noindent {\bf Case 1}: $\bar\varphi_\alpha=0, \forall\alpha$.
 \\
 \\
This reduces to the free-string case above, as here the off
diagonal terms of $M$ also vanish. We have a global minimum for
$\beta
> \beta_1$ as all the eigenvalues are positive.  Otherwise, when
$\beta_3 < \beta < \beta_1$ we have a saddle point, and for $\beta
< \beta_3$ a global maximum.

Thus, as the temperature increases (or $\beta$ decreases) the 1
direction will `roll' first.
\\
\\
\noindent {\bf Case 2}: $\bar\varphi_2 = \bar\varphi_3=0$ but
$\bar\varphi_1 \neq 0$.
 \\
 \\
Now, notice that the temperatures $\beta_1$, $\beta_2$ and
$\beta_3$, as defined for free strings, are in principle relevant
{\it only}  when $\bar\varphi_\alpha=0$ since then the
off-diagonal terms of $M$ vanish. When non-zero
$\bar\varphi_\alpha$ enter, we have to worry about the
off-diagonal terms, and find the new eigenvalues.  This in turn
will introduce new critical ($K$-dependent) temperatures.

As before, $\bar\varphi_1$ is the solution of the standard
equation {\it provided} $2D{\bar J}_1
>1$ or $\beta < \beta_1$.

 When $\beta=\beta_2$ the smallest eigenvalue is negative, showing
 that $({\bar\varphi}_1,0,0)$ is not a local minimum.
There is an intermediate temperature $\beta_*$, the solution to
 \be  (1-2DJ_2(\beta_*))(1-2DJ_3(\beta_*)) = K^2(\beta_*) u^2(\bar{\varphi}_1) \ee
 that denotes the transition from local minimum to saddle point.
 That is, strings of type 2
and 3 are nucleated at the same time.

To summarize: for $\beta > \beta_1$ there is a global minimum at
$\bar\varphi_\alpha=0$. For $\beta_2 < \beta_* < \beta < \beta_1$
the global minimum is at $(\bar{\varphi}_1,0,0)$.
\\
\\
\noindent {\bf Case 3}: $\bar\varphi_1,\bar\varphi_2,
\bar\varphi_3$ all non-zero.
 \\
 \\
For $\beta < \beta_*$ ,  type 2 and 3 strings are nucleated since
one cannot have only one non-zero $\bar \varphi_{\alpha}$. Hence we
expect to have non-zero $\bar\varphi_\alpha$ for all $\alpha$.
However, there is nothing at this stage to preclude the
possibility of even further transitions, of first and second
order.
\\
\\
\noindent {\bf Discussion}.
 \\
 \\
We can get some help from elementary Morse theory, applied to the
whole $\bar\varphi_{\alpha}$ space and not just the positive
sector \cite{Volovik}. Empirically, for the purpose of  counting
extrema, equations (\ref{11}) also behave just like the cubic
equations that would follow from taking only the leading terms
(\ref{V3}). According to this, when we have 27 extrema, no more
than 14 can be minima. The cases of all $\bar\varphi_{\alpha} = 0$
or one $\bar\varphi_{\alpha}$ non-zero may produce 7 ($7 = 1
+3\times 2$) real extrema and therefore 20 may correspond to
extrema with no $\bar\varphi_{\alpha}$ vanishing. From $D_2$, each
is fourfold degenerate, implying that there may exist five ($5 =
20/4$) different least symmetric extrema, of which no more than
three can be local minima. This still allows for either first or
second-order Hagedorn transitions as $\beta$ is reduced below
$\beta_*$ (or temperature increased).

\begin{figure}
\centerline{\includegraphics[width=0.4\textwidth]{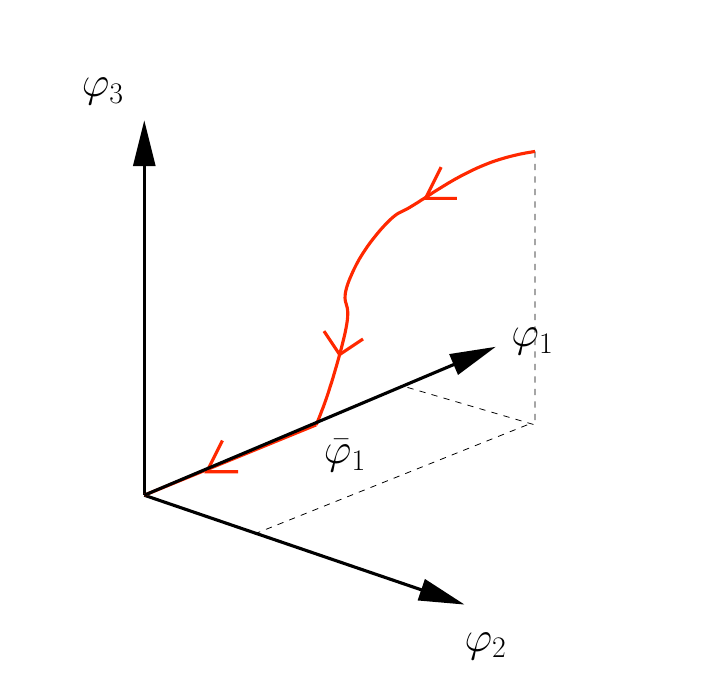}}
\caption{Schematic representation of the trajectory of $\varphi$
in field space.  The arrow indicates the trajectory as a function
of decreasing temperature.} \label{fig:2}
\end{figure}

Now consider the case when two string types have (approximately)
the same tension, and the other is markedly different, e.g.~one
string is very light, and the others heavy. The cases of all
$\bar\varphi_{\alpha} = 0$ or one $\bar\varphi_{\alpha}$ non-zero still
may produce 7 ($7 = 1 + 2+4$) real extrema. However, each extremum
with  no $\bar\varphi_{\alpha}$ vanishing is now approximately
eightfold symmetric. As a result we do not expect more than two of
them, of which only one can be a local minimum. This means that
there cannot be any further transitions as $\beta$ is reduced
below $\beta_*$. Although a first order transition cannot be
precluded, empirically we have only found second order transitions
even for $\sigma_{\alpha}$ taking different values.
The situation is summarised schematically in figure \ref{fig:2}.

From the above discussion, it follows trivially that, for equal $\sigma_{\alpha}$, (with
twelve-fold degeneracy for all $\varphi_{\alpha}$ non-zero) there
is just one second-order transition.
This is relevant to an idealised version of QCD. However, as it
stands the analysis above is restricted to closed or infinite
string. The addition of quarks to string ends changes the picture
again. Further, since flux tubes are not fundamental in any sense,
the `Hagedorn' transition in QCD has a different status, with no
ambiguity about increasing the temperature beyond it.
\\
\\
\noindent{\bf Density of vertices:}
 \\
 \\
Finally we end this section with a comment on the density of vertices in
the different phases.
From (\ref{nv}), and since $I_m(0) = 0$ for $m\geq 1$ it follows
that, on differentiating $V(\varphi_\alpha )$ with respect to $K$,
 \be
n_v = 0
 \ee
 when any ${\bar\varphi}_\alpha = 0$.
 Thus, we only have a non-zero density of vertices on infinite strings for
 $\beta<\beta_*$, i.e.~at temperatures high enough for all infinite
 string of all
 types to be present.  This is shown in Figure \ref{fig:1}.
 \\
 \\
\section{Conclusions}

The main idea of this paper has been very simple: that we can
describe the thermodynamics of a network of strings of three
different types (and tensions) by an effective three-field theory
whose potential $ V(\varphi_1,\varphi_2,\varphi_3)$ takes the form
 \ba
 \beta V = \sum_{\alpha}\bigg[
 \frac{1}{2}m_{\alpha}^2\varphi_{\alpha}^2 +
 \frac{1}{4}\lambda_{\alpha}\varphi_{\alpha}^4
 \bigg] +\mu\varphi_1\varphi_2\varphi_3 + ...\label{V3b}
 %\nn
 \ea
 The interaction coefficient $\mu$ reflects the presence of
 Y-junctions at which one string of each type meet.  The
 coefficients are temperature dependent, with $m_{\alpha}^2\propto (1-T/T_{\alpha})$
 in the vicinity of its zero. If $\mu$ were zero, the $T_{\alpha}$
 would be Hagedorn temperatures for the individual string types.
 As a result, the discrete symmetries of $V$ are {\it broken}
 at high temperature, {\it restored} at low temperature, in a reversal of the usual pattern.

Our main results, summarised Figs.~1 and 2, essentially follow
from the form of (\ref{V3b}) alone, supplemented by an
understanding of the order parameters, that they characterise
infinite string, and not loops. In consequence, in a network of
strings of different tensions it is the lightest strings whose
{\it infinite} strings survive last after Hagedorn transitions,
and even those disappear in turn, to leave a collection of small
loops. This is despite the presence of junctions between strings
of different types. That is, the only r\^ole that the junctions
play is in these small loops of string whose presence is the only
memory of the initial proliferation of strings of all types.

The burden of this paper has been  to provide a model in which we
can see how the potential (\ref{V3b}) is realised, almost as proof
of principle. This has turned out to be a non-trivial task and the
model at hand, an extension of similar models used in QCD in a
much more restricted situation, has its faults.  A well as picking
a path through the `fermionic' lattice artefacts, as in the
calculations for QCD strings, our strings are also assumed to be
non-interacting and static. Furthermore we are often pushed to
consider the model in a limit of parameter space where
approximations are not always well controlled (just as in
\cite{Patel1,Patel2,MR}).  Our one string bosonic model
demonstrated how, for a single field, $\mu\varphi^3$ terms arise
naturally. However, being unable to generalise the bosonic model
to three string types, we have also had to introduce massive
vertices in the three string model as an artefact of the lattice
mean-field approximation. Naturally, any specific model will give
more information than just the leading terms of $V$ of
(\ref{V3b}). In our case the model is a generalised XY model, in
which transitions are seen in the language of spin ordering and
which, in principle, permit better than the mean-field
approximation.

As suggested above, our analysis points to the the final stage of
the transitions as being that of a single string type, collapsing
into loops, which was the original case to be studied, primarily
in the context of Nambu-Goto strings. In that case, the full
statistical mechanics has been studied in detail, and can be
generalised to non-static strings.  The result however, is the
same!  Indeed, rather than consider random walks in space, one can
consider simultaneous independent random walks on the
 Kibble-Turok spheres for left and right-moving modes respectively \cite{AlbrechtTurok}. The
microstate density at the transition is the square of that for simple random walks,
but integrating over centre-of-mass coordinates reduces the state
density to that of (appropriately defined) single static random
walks.

Another way to make this adiabatic picture dynamical is to attempt
to determine the time scales of the string network transitions
from the timescales of the effective field theory, using the
Kibble scenario \cite{Kibble}. This relies on little more than
causal bounds, and the analysis is under way.

\section*{Acknowledgements}
We thank Ed Copeland, Mark Hindmarsh, Tom Kibble and Mairi
Sakellariadou for useful discussions. RJR thanks the CNRS and the University of Paris 7 for
financial support, and is grateful to APC, Paris 7, and the LPT in Orsay for warm
hospitality.

%\end{widetext}

\end{document}